\documentclass[a4paper]{jpconf}
\usepackage{graphicx}
\usepackage{amsmath}
\newcommand{\mgm}{\ensuremath{\mu}}
\begin{document}
\title{Status and perspectives of neutrino magnetic moments}

\author{Alexander Studenikin}

\address{Department of Theoretical Physics, Faculty of Physics, Lomonosov Moscow State University, Moscow 119991,
Russia\\
Joint Institute for Nuclear Research, Dubna 141980, Moscow
Region,
Russia}
\ead{studenik@srd.sinp.msu.ru}

\begin{abstract}
Basic theoretical and experimental aspects of neutrino magnetic moments are reviewed, including the present best upper bounds from reactor experiments and astrophysics. An interesting effect of neutrino spin precession and oscillations induced by the background matter transversal current or polarization is also discussed.
\end{abstract}

\section{Introduction}
The period of about last one and a half decade has been celebrated by distinguished discoveries in particle physics that were honored by two Nobel Prizes. The Nobel Prize in 2013 was awarded jointly to Francois Englert and Peter Higgs for the theoretical discovery of a mechanism that contributes to our understanding of the origin of mass of subatomic particles, and which recently was confirmed through the
discovery of the Higgs boson at LHC. This provides the final glorious triumph of the Standard Model. The Nobel Prize of 2015 awarded to Arthur McDonald and Takaaki  Kajita for the discovery of neutrino oscillations, which shows that neutrinos have mass.
In the Standard Model neutrinos are massless particles. Thus, our present knowledge on properties of neutrinos opens a window to new physics beyond the Standard Model.

In many extensions of the Standard Model, which account for neutrino masses and mixings, neutrinos acquire nontrivial electromagnetic properties that hence allow direct electromagnetic interactions of neutrinos with electromagnetic fields and other particles. The most well studied and understood among the neutrino electromagnetic characteristics are the magnetic moments. A complete recent review on the neutrino electromagnetic properties with a detailed list of references to the corresponding literature can be found in \cite{Giunti:2014ixa}.

In this note, after a short introduction devoted to neutrino electromagnetic vertex and to neutrino magnetic moments in gauge models, the studies of neutrino magnetic moments in the reactor neutrino experiments are discussed. After that we flash up the neutrino electromagnetic interactions in astrophysics and the astrophysical bounds on neutrino magnetic moments. Then we discuss  the most recent new studies and prospects in neutrino magnetic moments that have appeared after publication of our review paper \cite{Giunti:2014ixa}. The final section of the paper is devoted to an interesting effect of neutrino spin precession and oscillations induced by the background matter transversal current or polarization. This effect was proposed and pronouncedly
discussed in our papers \cite{Studenikin:2004bu, Studenikin:2004tv}
 and has been recently confirmed and used in
\cite{Kartavtsev:2015eva,Volpe:2015rla,Cirigliano:2014aoa}
for an analysis of neutrinos propagation in anisotropic astrophysical environments.

\section{Neutrino electromagnetic vertex}

The neutrino electromagnetic properties are determined by the neutrino electromagnetic vertex
function $\Lambda_{\mu}(q)$
that is related to the matrix element of the electromagnetic current between the
neutrino initial $\psi (p)$ and final $\psi (p')$ mass states
\begin{equation}\label{matr_elem}
<{\psi}(p^{\prime})|J_{\mu}^{EM}|\psi(p)>= {\bar
u}(p^{\prime})\Lambda_{\mu}(q)u(p).
\end{equation}
 The Lorentz and electromagnetic gauge invariance imply
 \cite{Giunti:2014ixa,Kayser:1982br,Nieves:1981zt,
Nowakowski:2004cv}
that the electromagnetic vertex function can be written in the form:
\begin{equation}
\Lambda_{\mu}(q)
=
f_{Q}(q^{2}) \gamma_{\mu}
-
f_{M}(q^{2}) i \sigma_{\mu\nu} q^{\nu}
+
f_{E}(q^{2}) \sigma_{\mu\nu} q^{\nu} \gamma_{5}
+
f_{A}(q^{2}) (q^{2} \gamma_{\mu} - q_{\mu} {q}) \gamma_{5}
,
\label{vert_func}
\end{equation}
where
$f_{Q} $,
$f_{M} $,
$f_{E} $ and
$f_{A} $ are charge, dipole magnetic and electric and anapole neutrino electromagnetic form factors, and $\sigma_{\mu\nu}=\frac{i}{2}(\gamma _{\mu}\gamma_{\nu}-\gamma _{\nu}\gamma_{\mu})$.
The photon four-momentum  $q$ is given by $q=p-p'$ and form factors depend on the Lorentz invariant dynamical quantity $q^2$.

Note that the Dirac and Majorana neutrinos have quite different electromagnetic properties. This can be easily seen \cite{Giunti:2014ixa} if one accounts for constraints on form factors imposed by the the hermiticity of the neutrino electromagnetic current and its invariance under discrete symmetries transformations. In the case of Dirac neutrinos, the assumption of $CP$ invariance combined with the hermiticity of $J_{\mu}^{EM}$ implies that $f_{E}(q^2)=0$. In the case of Majorana neutrinos from the $CPT$ invariance, regardless of whether $CP$ invariance is violated or not, $f_Q(q^2)=f_M(q^2)=f_E(q^2)=0$ and only the anapole form factor $f_{A}(q^2)$ can be nonvanishing.

In general case the matrix element of the electromagnetic current (\ref{matr_elem}) can be considered between neutrino initial $\psi (p)$ and final $\psi (p')$ states of different masses $p^2 = m_i ^2\neq m_j ^2={p'}^2$. Then the vertex function and form factors are matrixes in the neutrino mass eigenstates space,
\begin{equation}
\label{Lambda} \Lambda_{\mu}(q)_{ij}=
\Big(f_{Q}(q^{2})_{ij}+f_{A}(q^{2})_{ij}\gamma_{5}\Big)
(q^{2}\gamma_{\mu}-q_{\mu}{\not
q})+ f_{M}(q^{2})_{ij}i\sigma_{\mu\nu}q^{\nu}
+f_{E}(q^{2})_{ij}\sigma_{\mu\nu}q^{\nu}\gamma_{5}.
\end{equation}
For two Dirac neutrinos in the off-diagonal case ($i\neq j$) the hermiticity by itself does not
imply restrictions on the form factors and all of them can be nonzero.
Two massive Majorana neutrinos can have either
a transition electric form factor
or a transition magnetic form factor,
but not both, and the transition electric form factor can exist only together with a transition anapole form factor, whereas the transition magnetic form factor can exist only together with a transition charge form factor.

 From the demand that the form factors at
zero momentum transfer, $q^2=0$, are elements of the scattering
matrix, it follows that  in any consistent theoretical model the form
factors in the matrix element (\ref{matr_elem}) should be gauge
independent and finite. Then, the form factors values at $q^{2}=0$
determine the static electromagnetic properties of the neutrino that
can be probed or measured in the direct interaction with external
electromagnetic fields.

\section{Magnetic  dipole moment in gauge models}

The most well studied and understood among the neutrino electromagnetic characteristics are the dipole magnetic (diagonal, $i=j$, and transition, $i\neq j$) moments
\begin{equation}
\mu_{ij}=f_{M}(0)_{ij},
\end{equation}
given by the corresponding form factors  at $q^2=0$.

The diagonal magnetic  moment of a Dirac neutrino in the minimally-extended Standard Model with right-handed neutrinos, derived for the first time in \cite{Fujikawa:1980yx}, is respectively,
\begin{equation}\label{mu_D}
    \mu^{D}_{ii}
  = \frac{3e G_F m_{i}}{8\sqrt {2} \pi ^2}\approx 3.2\times 10^{-19}
  \Big(\frac{m_i}{1 \ eV }\Big) \mu_{B}
\end{equation}
where $\mu_B$ is the Borh magneton.
Note that the estimation (\ref{mu_D}) of the obtained value for the neutrino magnetic moment  shows that it is very small. It is indeed very small if compare, for instance, with the similar characteristic of electromagnetic properties of  charged leptons ($l=e, \mu, \tau $) that are the anomalous magnetic moments given by
\begin{equation}\label{AMM}
\mu^{AMM}_{l}=\frac{\alpha_{QED}}{2\pi}\mu_{B} \sim 10^{-3} \mu_{B}.
\end{equation}
There is also a reasonable gap between the prediction (\ref{mu_D}) of the minimally-extended Standard Model with right-handed neutrinos and the present experimental and astrophysical upper bounds on the neutrino effective magnetic moments. However, in many other theoretical frameworks beyond the minimally-extended Standard Model the neutrino magnetic moment can reach values being of interest for the next generation terrestrial experiments and also accessible for astrophysical observations.

The transition magnetic moment of a Dirac
neutrino is given by \cite{Shrock:1982sc,Pal:1981rm,Bilenky:1987ty}
\begin{equation}\label{mu_D_epsilon_D_trans}
    \mu^{D}_{ij}=\frac{3e G_F m_{i}}{32\sqrt {2} \pi ^2}
  \Big(1+ \frac{m_j}{m_i}\Big)
  \sum_{l=\ e, \ \mu, \ \tau}\Big(\frac{m_l}{m_W}\Big)^2U_{lj}U^{\ast}_{li}.
\end{equation}
Numerically  transition moments of a Dirac neutrino
can be expressed as follows
\begin{equation}
    \mu^D_{ij}=4\times 10^{-23} \mu_{B}\Big(\frac{m_i + m_j}{1\ {eV}}\Big)
  \sum_{l=\ e, \ \mu, \ \tau}\Big(\frac{m_l}{m_\tau}\Big)^2U_{lj}U^{\ast}_{li}.
\end{equation}
The transition magnetic moment $\mu^{D}_{ij}$ is even much smaller than the diagonal magnetic moment $\mu^{D}_{ii}$ because of the leptonic GIM mechanism. That is a reason why the neutrino radiative decay $\nu_i\rightarrow \nu_j +\gamma$ is in general a very slow process.

The transition magnetic moments of a Majorana neutrino are given by
\cite{Shrock:1982sc}
\begin{equation}\label{mu_M_trans}
\mu^{M}_{ij}
=-
\frac{3e G_F m_{i}}{16\sqrt{2}\pi^{2}}
\left( 1 + \frac{m_{j}}{m_{i}} \right)
\sum_{l=e,\mu,\tau}
{Im}\left[U^{*}_{lk} U_{lj}\right]
\,
\frac{m_{l}^{2}}{m_{W}^{2}}.
\end{equation}
There is the increase by a factor of 2 of the first coefficient
with respect to the Dirac case in (\ref{mu_D_epsilon_D_trans})
\cite{Schechter:1981hw, Pal:1981rm}.

The dependence of the diagonal
magnetic moment of a massive Dirac neutrino on the neutrino $b_{i}=m_{i}^{2}/m_{W}^{2}$
and charged lepton $a_{l}=m_{l}^{2}/m_{W}^{2}$ mass parameters in the one-loop approximation in the minimally-extended Standard Model with right-handed neutrinos was studied in  \cite{CabralRosetti:1999ad,Dvornikov:2003js,Dvornikov:2004sj}.
The calculations of the neutrino magnetic
moment which take into account exactly the dependence on the
masses of all particles \cite{Dvornikov:2003js,Dvornikov:2004sj} can be useful
in the case of existence of a heavy neutrino with a mass comparable or even exceeding the values of
the masses of other known particles.



As it has been already mentioned, the linear dependence of the neutrino magnetic moment on the neutrino mass makes its value in general very small. This feature is common for a vide class of theoretical models and seems to be hardly avoidable. There is also indeed a general problem for a theoretical model of how to get a large magnetic
moment for a neutrino and simultaneously to avoid an unacceptable large
contribution to the neutrino mass. If a contribution to the neutrino
magnetic moment of order $\mu_{\nu} \sim \frac{eG}{\Lambda}$
is generated by physics beyond the minimally-extended Standard Model at an
energy scale characterized by $\Lambda$, then the correspondent
contribution to the neutrino mass is \cite{Pal:1991pm,Bell:2005kz,Balantekin:2006sw}
\begin{equation}\label{mu_Lambda}
\delta m_{\nu} \sim \frac{\Lambda ^2}{2m_e}\frac{\mu_{\nu}}{\mu_B}=
\frac{\mu_{\nu}}{10^{-18}\mu_B}\Big(\frac{\Lambda}{1 \ Tev}\Big)^2\
eV.
\end{equation}
Therefore, a particular fine tuning is needed to get a large value
for the neutrino magnetic moment while keeping the neutrino mass
within experimental bounds. Different possibilities to have a large
magnetic moment for a neutrino were considered
 in the literature starting with \cite{Voloshin:1987qy,Barr:1990um}, one of the possibilities
have been considered recently \cite{Xing:2012gd}.

\section{Neutrino electromagnetic properties in reactor experiments}

The most established and sensitive
method for the experimental investigation of neutrino electromagnetic properties
is provided by direct laboratory measurements of
antineutrino-electron scattering at low energies in solar,
accelerator and reactor experiments. A detailed description of
different experiments can be found in~
\cite{Giunti:2014ixa, Balantekin:2006sw, Wong:2005pa,  Beda:2007hf,Giunti:2008ve,
Broggini:2012df}.
Here below we focus on the reactor antineutrino experiment since it provides the best nowadays terrestrial laboratory upper limit on the effective neutrino magnetic moment, as well a limit
on the neutrino millicharge \cite{Studenikin:2013my} and probably with the expected improvement of sensitivity will be also sensitive in future to the neutrino charge radius \cite{Kouzakov:2015}.

In general case the cross section for an antineutrino scattering on a free
electron can be written~\cite{Vogel:1989iv} as a sum of the Standard Model weak interaction and the electromagnetic interaction
contributions,
\begin{equation}
\label{cr_sec}\frac{d\sigma}{dT}=\Big(\frac{d\sigma}{dT}\Big)_{SM}+
\Big(\frac{d\sigma}{dT}\Big)_{EM}.
\end{equation}

The Standard Model contribution is
\begin{equation}\label{d_sigma_SM}
\Big(\frac{d\sigma}{dT}\Big)_{SM}=\frac{G^2_F m_e}{2\pi}\Bigg[(g_V +
g_A)^2 + (g_V - g_A)^2\Big(1-\frac{T}{E_\nu}\Big)^2
 +(g_A^2 -
g_V^2)\frac{m_eT}{E^2_\nu}\Bigg],
\end{equation}
where $E_\nu$ is the initial neutrino energy and $T$ is the electron
recoil energy which is measured in the experiment. The coupling
constants are $g_V={2\sin ^2 \theta _W + \frac{1}{2}}$ and $g_A=-\frac{1}{2}$.
In the case $E_\nu\gg T$, which
is relevant to the experiments with reactor antineutrinos
\begin{equation} \label{SM}
\Big(\frac{d\sigma}{dT}\Big)_{SM}= {G_F^2  m \over 2 \pi} \left( 1+ 4 \sin^2
\theta_W + 8 \sin^4 \theta_W \right) \left [ 1 + O \left ( {T
\over E_\nu} \right) \right ].
\end{equation}

The electromagnetic interaction part of the cross section can be written as a sum of three contributions originated due to the neutrino magnetic moment, millicharge and charge radius respectively,
\begin{equation}\label{d_sigma_EM}
\Big(\frac{d\sigma}{dT}\Big)_{EM}=\Big(\frac{d\sigma}{dT}\Big)_{\mu_{\nu}}+\Big(\frac{d\sigma}{dT}\Big)_{q_{\nu}}+
\Big(\frac{d\sigma}{dT}\Big)_{\langle r_{\nu_e}^2\rangle}.
\end{equation}
The contribution to the cross section generated by possible non-zero charge radius $\langle r_{\nu_e}^2\rangle$ is given by (\ref{d_sigma_SM}) with the redefined vector coupling constant \cite{Vogel:1989iv}
$g_V\rightarrow \frac{2}{3}m^2_W {\langle r_{\nu_e}^2\rangle}\sin^2 \theta_W$,
\begin{equation}\label{d_sigma_ch_rad}
\Big(\frac{d\sigma}{dT}\Big)_{\langle r_{\nu_e}^2\rangle}=\Big(\frac{d\sigma}{dT}\Big)_{{EM}_{{g_V\rightarrow \frac{2}{3}m^2_W {\langle r_{\nu_e}^2\rangle}\sin^2 \theta_W}}}.
\end{equation}
The magnetic moment contribution to the cross section and its expression at $E_\nu \gg T$ are as follows,
\begin{equation} \label{d_sigma_mu}
\Big(\frac{d\sigma}{dT}\Big)_{\mu_{\nu}}=  \pi {\alpha_{QED}^2 \over m_e^2} \left ( {\mu_\nu \over \mu_B} \right )^2
\left ( {1 \over T} - {1 \over E_\nu } \right )\approx
\pi\alpha_{QED}^{2}\frac{1}{m_{e}^{2}T}
\left(\frac{\mu_{\nu}}{\mu_{B}}\right)^{2}.
\end{equation}
The magnetic moment contribution to the cross section also changes the helicity of the neutrino, contrary to all other discussed contributions. The contribution to the cross section due to the neutrino millicharge is
\begin{equation}\label{sigma_q_e}
\left(\frac{d\sigma}{dT}\right)_{q_{\nu}}\approx 2\pi\alpha
\frac{1}{m_{e}T^2}q_{\nu}^2.
\end{equation}
Note that three terms $\Big(\frac{d\sigma}{dT}\Big)_{SM}$, \ $\Big(\frac{d\sigma}{dT}\Big)_{\mu_{\nu}}$ and $\Big(\frac{d\sigma}{dT}\Big)_{q_{\nu}}$ exhibit quite different dependence on the electron recoil energy $T$.

 The strategy of the experiment is to find an excess of events over those due to the Standard Model and other background processes. Experimental
signatures for $\mu_{\nu}$ or $q_{\nu}$ would be an excess of events between the reactor ON over OFF
samples, which exhibits an $\frac{1}{T}$ or $\frac{1}{T^2}$ energy dependence.  It is clear that the lower the measured electron recoil energy $T$ is the smaller neutrino magnetic moment $\mu_{\nu}$ and millicharge $q_{\nu}$ values can be probed in the experiment.

Up to now in the reactor experiments there is no any access of events due to possible neutrino electromagnetic interactions.
From comparison of two cross sections (\ref{SM}) and (\ref{d_sigma_mu})
the sensitivity of the experiment to the neutrino magnetic moment $\mu^2_\nu$ can be estimated:
\begin{equation}\label{mu_sensitivoty}
\mu^2_\nu\leq \frac{G_F^2 m^3_e T}{2\pi^2\alpha_{QED}^2}(1+4\sin^2_W+8\sin^2_W)\mu_B ^2.
\end{equation}
The best laboratory upper limit on a neutrino magnetic moment \cite{Beda:2012zz} has been obtained by the GEMMA collaboration  (Germanium Experiment for measurement of Magnetic Moment of Antineutrino) that investigates the reactor antineutrino-electron scattering at the Kalinin Nuclear Power Plant (Russia). Within the presently reached electron recoil energy threshold of
$T \sim 2.8 \ keV$
the neutrino magnetic moment is bounded from above by the value
\begin{equation}\label{mu_bound}
\mu_{\nu} < 2.9 \times 10^{-11} \mu_{B} \ \ (90\% \ C.L.).
\end{equation}
This limit obtained from unobservant distortions in the recoil electron energy
spectra is valid for both Dirac and Majorana neutrinos and for both diagonal
and transitional moments. The next edition of the antineutrino scattering experiment at the Kalinin Nuclear Power Plant will provide a reasonable improvement of upper bounds on neutrino magnetic
moments (for the corresponding estimations see in \cite{Beda:2012zz,Beda:2013mta}).

Stringent upper bounds on the neutrino magnetic moment have been also
obtained in other recently carried reactor experiments: $\mu_\nu \leq
9.0 \times 10^{-11}\mu_{B}$ (MUNU collaboration \cite{Daraktchieva:2005kn})
and $\mu_\nu \leq 7.4 \times 10^{-11}\mu_{B}$ (TEXONO collaboration
\cite{Wong:2006nx}). Stringent limits  also
obtained  in the solar neutrino scattering experiments: $\mu_\nu \leq
1.1 \times 10^{-10}\mu_{B}$ (Super-Kamiokande collaboration
\cite{Liu:2004ny}) and $\mu_\nu \leq 5.4 \times 10^{-11}\mu_{B}$
(Borexino collaboration \cite{Arpesella:2008mt}).
Quite recent  new analysis \cite{Canas:2015yoa} of the Borexino data
yields more stringent bound   $\mu_\nu \leq 3.1 \times 10^{-11}\mu_{B}$.

Interpretations and comparisons among various experiments should take into account the difference in the
flavour compositions between them at the detectors.
It should be mentioned \cite{Beacom:1999wx} that what is measured in scattering
experiments is an effective magnetic moment $\mu^{exp}_{l}$, that
depends on the flavour composition of the neutrino beam at the
detector located at a distance $L$ from the source, and which value
is a rather complicated function of the magnetic (transition) moments
$\mu_{i j}$:
\begin{equation}\label{mu_exp}\nonumber
{\mu^{exp}_{\nu_l}} ^{2} = \mu_{\nu}^{2}(\nu_{l},L,E_{\nu})=\sum_{j}\Big|
\sum_{i} U_{li} e^{-iE_{i}L}(\mu_{ji}-i\epsilon_{ji}) \Big| ^{2}.
\end{equation}
The dipole electric (transition) moments, if these quantities not
vanish, can also contribute to $\mu^{exp}_{l}$.  For completeness we mention that
the absence of distortions of the electron recoil energy spectra in GEMMA experiment
was used in \cite{Studenikin:2013my} to bound from above the neutrino millicharge.
The possibility to constrain $q_{\nu}$ with germanium detectors at sub-keV
sensitivities have been recently considered also in \cite{Chen:2014dsa}.
The most stringent astrophysical constraint on $q_{\nu}$ has been recently obtained
in \cite{Studenikin:2012vi}.

\section{Recent studies and prospects in neutrino magnetic moments}

It is planned \footnote {Victor Brudanin and Vyacheslav Egorov,
private communication.} that a new experiment at the Kalinin Nuclear Power Plant, which is now  in the final stage of preparation and
is expected to get data within a period of one year, the effective electron
threshold will be reduced to $T= 350 \ eV$ and the sensitivity to the neutrino magnetic moment will be at the level
$\mu_{\nu}\sim 9\times 10^{-12} \mu_{B}$.

A new possibility to distinguish the Dirac and Majorana nature of neutrinos comparing the values of effective neutrino magnetic moments $\mu_{\nu_e}, \mu_{\nu_\mu}, \mu_{\nu_\tau}$, if measured in corresponding experiments, is proposed in \cite{Frere:2015pma}. It is shown that in the case of Majorana neutrinos triangle inequalities, $ \mid \mu_{\nu_\tau}\mid ^2 \leq \mid \mu_{\nu_\e}\mid ^2 + \mid \mu_{\nu_\mu}\mid ^2$ and
cyclic permutations, have to fulfill, which do not hold for Dirac neutrinos. Observing a violation
of these inequalities would prove either at the Dirac nature of neutrinos or at the presence of an extra light sterile neutrino mode.

\section{Neutrino electromagnetic interactions in astrophysics}

If a neutrino has
the non-trivial electromagnetic properties discussed above, a direct
neutrino coupling to photons is possible and several processes important
for applications in astrophysics exist \cite{Raffelt:1996wa}. A set of
most important neutrino electromagnetic processes is: 1) neutrino
radiative decay $\nu_{1}\rightarrow \nu_{2} +\gamma$, neutrino
Cherenkov radiation in an external environment (plasma and/or
electromagnetic fields), spin light of neutrino, $SL\nu$ , in the
presence of a medium \cite{Lobanov:2002ur,Studenikin:2004dx,Grigorev:2005sw,Grigoriev:2012pw}; \ 2) photon (plasmon) decay to a
neutrino-antineutrino pair in plasma $\gamma \rightarrow \nu {\bar
\nu }$; \ 3) neutrino scattering off electrons (or nuclei); \ 4)
neutrino spin (spin-flavor) precession in a magnetic field (see
\cite{Okun:1986na}) and resonant neutrino spin-flavour
oscillations in matter \cite{Lim:1987tk,Akhmedov:1988uk}.

The tightest astrophysical bound on a neutrino magnetic moment is
provided by observed properties of globular cluster stars. For a
large enough neutrino magnetic moment the plasmon decay rate can be
enhanced so that a reasonable delay of helium ignition would appear.
From lack observation evidence of anomalous stellar cooling due to
the plasmon decay the following limit has been found \cite{Raffelt:1990pj}
\begin{equation}
\Big( \sum _{i,j}\mid \mu_{ij}\mid ^2\Big) ^{1/2}\leq 3 \times
10^{-12} \mu _B.
\end{equation}
This is the most stringent astrophysical constraint on neutrino
magnetic moments, applicable to both the Dirac and Majorana neutrinos.

%
\section{An effect of neutrino spin precession due to matter motion}

An approach based on the generalized Bargmann-Michel-Telegdi equation can be used for derivation of an impact of matter motion and polarization on the neutrino spin (and spin-flavour) evolution. Consider, as an example, an electron neutrino spin procession in the case when neutrinos with the Standard Model interaction are propagating through moving and polarized matter composed of electrons (electron gas) in the presence of an electromagnetic field given by the electromagnetic-field tensor $F_{\mu \nu}=({\vec E}, {\vec B})$. As discussed in \cite{Studenikin:2004bu, Studenikin:2004tv}
(see also \cite{Egorov:1999ah,Lobanov:2001ar, Dvornikov:2002rs}) the evolution of the
three-di\-men\-sio\-nal neutrino spin vector $\vec S $ is given by
\begin{equation}\label{S} {d\vec S
\over dt}={2\mgm\over \gamma} \Big[ {\vec S \times ({\vec
B_0}+\vec M_0)} \Big],
\end{equation}
where the magnetic field $\vec{B_0}$ in the neutrino rest frame is determined by the transversal
and longitudinal (with respect to the neutrino motion) magnetic and electric field components in the
laboratory frame,
\begin{equation}
\vec B_0=\gamma\Big(\vec B_{\perp} +{1 \over \gamma} \vec
B_{\parallel} + \sqrt{1-\gamma^{-2}} \Big[\vec E_{\perp} \times
\frac{\vec \beta}{\beta} \Big]\Big),
\end{equation}
$\gamma = (1-\beta^2)^{-{1 \over 2}}$, $\vec \beta$ is the neutrino velocity.

The matter term $\vec M_0$ in Eq. (\ref{S}) is also composed of the transversal $\vec {M}{_{0_{\parallel}}}$
and longitudinal  $\vec {M_{0_{\perp}}}$ parts,
\begin{equation}
\vec {M_0}=\vec {M}{_{0_{\parallel}}}+\vec {M_{0_{\perp}}},
\label{M_0}
\end{equation}
\begin{equation}
\begin{array}{c}
\displaystyle \vec {M}_{0_{\parallel}}=\gamma\vec\beta{n_{0} \over
\sqrt {1- v_{e}^{2}}}\left\{ \rho^{(1)}_{e}\left(1-{{\vec v}_e
\vec\beta \over {1- {\gamma^{-2}}}} \right)\right.-
\displaystyle{\rho^{(2)}_{e}\over {1- {\gamma^{-2}}}}\left. \left(\vec\zeta_{e}\vec\beta
\sqrt{1-v^2_e}+ {(\vec \zeta_{e}{\vec v}_e)(\vec\beta{\vec v}_e)
\over 1+\sqrt{1-v^2_e} }\right)
\right\}, \label{M_0_parallel}
\end{array}
\end{equation}
\begin{equation}\label{M_0_perp}
\begin{array}{c}
\displaystyle \vec {M}_{0_{\perp}}=-\frac{n_{0}}{\sqrt {1-
v_{e}^{2}}}\Bigg\{ \vec{v}_{e_{\perp}}\Big(
\rho^{(1)}_{e}+\rho^{(2)}_{e}\frac
{(\vec{\zeta}_{e}{\vec{v}_e})} {1+\sqrt{1-v^2_e}}\Big) +
\displaystyle {\vec{\zeta}_{e_{\perp}}}\rho^{(2)}_{e}\sqrt{1-v^2_e}\Bigg\}.
\end{array}
\end{equation}
Here $n_0=n_{e}\sqrt {1-v^{2}_{e}}$ is the invariant number density of
matter given in the reference frame for which the total speed of
matter is zero. The vectors $\vec v_e$, and $\vec \zeta_e \
(0\leq |\vec \zeta_e |^2 \leq 1)$ denote, respectively,
the speed of the reference frame in which the mean momentum of
matter (electrons) is zero, and the mean value of the polarization
vector of the background electrons in the above mentioned
reference frame. The coefficients $\rho^{(1,2)}_e$ calculated
within the extended Standard Model supplied with $SU(2)$-singlet right-handed neutrino
$\nu_{R}$ are respectively,  $\rho^{(1)}_e={\tilde{G}_F \over {2\sqrt{2}\mu }}, \ \ \rho^{(2)}_e =-{G_F \over {2\sqrt{2}\mu}}$,
where $\tilde{G}_{F}={G}_{F}(1+4\sin^2 \theta _W).$
For neutrino evolution between two neutrino states $\nu_{e}^{L}\Leftrightarrow\nu_{e}^{R}$ in presence of the magnetic field and moving matter we get the following equation
\begin{equation}\label{2_evol_eq}
	i\frac{d}{dt} \begin{pmatrix}\nu_{e}^{L} \\ \nu_{e}^{R} \\  \end{pmatrix}={\mgm}
	\begin{pmatrix}
	{1 \over \gamma}\big|{\vec
M}_{0\parallel}+{\vec B}_{0\parallel}\big| & \big|{\vec B}_{\perp} + {1\over
\gamma}{\vec M}_{0\perp} \big|  \\
	 \big|{\vec B}_{\perp} + {1\over
\gamma}{\vec M}_{0\perp} \big| & -{1 \over \gamma}\mid{\vec
M}_{0\parallel}+{\vec B}_{0\parallel}\big|  \\
		\end{pmatrix}
	\begin{pmatrix}\nu_{e}^{L} \\ \nu_{e}^{R} \\ \end{pmatrix}.
\end{equation}
Thus, the probability of the neutrino spin oscillations in the adiabatic
approximation is given by (see \cite{Studenikin:2004bu, Studenikin:2004tv})
\begin{equation}\label{ver2}
P_{\nu_L \rightarrow \nu_R} (x)=\sin^{2} 2\theta_\textmd{eff}
\sin^{2}{\pi x \over L_\textmd{eff}},\ \sin^{2} 2\theta_\textmd{eff}={E^2_\textmd{eff} \over
{E^{2}_\textmd{eff}+\Delta^{2}_\textmd{eff}}}, \ \ \
L_\textmd{eff}={2\pi \over
\sqrt{E^{2}_\textmd{eff}+\Delta^{2}_\textmd{eff}}},
\end{equation}
where
\begin{equation}\label{E3}
E_\textmd{eff}=\mgm \big|{\vec B}_{\perp} + {1\over
\gamma}{\vec M}_{0\perp} \big|, \ \
\Delta_ \textmd{eff}={\mgm \over \gamma}\big|{\vec
M}_{0\parallel}+{\vec B}_{0\parallel} \big|.
\end{equation}
It follows that even without presence of an electromagnetic field,
${\vec B}_{\perp}={\vec B}_{0\parallel}=0$,
neutrino spin  oscillations can be induced in the presence of matter
when the transverse matter term ${\vec M}_{0\perp}$ is not zero.
This possibility is realized in the case when the transverse component of the background matter velocity or its transverse polarization is not zero. It is obvious that for neutrinos
with nonzero transition magnetic moments a similar effect of spin-flavour
oscillations exists under the same background conditions. A possibility of neutrino
spin procession and oscillations induced by the transversal matter current or polarization
was first discussed in \cite{Studenikin:2004bu, Studenikin:2004tv}.
The existence of this effect has been recently confirmed  in
\cite{Kartavtsev:2015eva,Volpe:2015rla,Cirigliano:2014aoa} where neutrinos
propagation in anisotropic media is studied.

The author is thankful to Nicolao Fornengo and Carlo Giunti for the kind
invitation to participate in the 14th International Conference on Topics in Astroparticle
and Underground Physics. The work on this paper was partially supported by the Russian Basic Research Foundation grants no. 14-22-03043-ofi-m, 15-52-53112-gfen and 16-02-01023-a.

\end{document}